\documentclass[sigconf]{acmart}
\AtBeginDocument{%
  }

\setcopyright{none}
\copyrightyear{2026}
\acmYear{2026}
\acmConference[COLIEE 2026]{Workshop on the Thirteenth International Competition on Legal Information Extraction and Entailment}{June 12, 2026}{Singapore}

\usepackage{booktabs}
\usepackage{threeparttable}
\usepackage{graphicx}
\usepackage{multirow}
\usepackage{tikz}
\usetikzlibrary{positioning, arrows.meta, fit, backgrounds}

\newcommand{\fragmentsuppressedmarker}{\texttt{<FRAGMENT\_\allowbreak{}SUPPRESSED>}}

\renewcommand\footnotetextcopyrightpermission[1]{}
\begin{document}

\title[ABAI at COLIEE 2026 Task 1]{ABAI at COLIEE 2026 Task 1: Multi-Stage Retrieval with GraphRAG-Enhanced Meta-Learning, and a Post-Hoc Study of the Cross-Validation-to-Test Gap}

\author{Minhan Cho}
\authornote{Both authors contributed equally to this research.}
\email{zvezda@g.skku.edu}
\affiliation{%
  \institution{AlphaBridge}
  \city{Seongnam-si}
  \state{Gyeonggi-do}
  \country{Republic of Korea}
}
\affiliation{%
  \institution{Sungkyunkwan University}
  \city{Seoul}
  \country{Republic of Korea}
}
\author{Soyoung Park}
\authornotemark[1]
\email{attorney.soyoung@gmail.com}
\affiliation{%
  \institution{National Assembly Research Service}
  \city{Seoul}
  \country{Republic of Korea}
}
\affiliation{%
  \institution{Sungkyunkwan University}
  \city{Seoul}
  \country{Republic of Korea}
}
\author{Daejin Choi}
\authornote{Corresponding author.}
\email{djchoi@ewha.ac.kr}
\affiliation{%
  \institution{Ewha Womans University}
  \city{Seoul}
  \country{Republic of Korea}
}
\author{Jinyoung Han}
\authornote{Corresponding author.}
\email{jinyounghan@skku.edu}
\affiliation{%
  \institution{Sungkyunkwan University}
  \city{Seoul}
  \country{Republic of Korea}
}

\renewcommand{\shortauthors}{Cho et al.}

\begin{abstract}
  We present the ABAI submission to COLIEE 2026 Task~1, case law retrieval, together with a controlled study of why it underperformed.
  The task suppresses the cited passages themselves, which removes much of the lexical overlap a retriever would rely on.
  Our pipeline answers this with four independently trained stages: multi-view BM25 over citation-context windows with reciprocal rank fusion, neural reranking, graph-based features from entity communities and a graph attention network, and a LightGBM meta-learner over 34 features.
  Our best run reached F1=0.177 on the official test set, against a cross-validated 0.311, and we attributed that gap to a recall ceiling, temporal distribution shift, and threshold miscalibration.
  We then tested all three.
  Under leakage-free protocols threshold transfer costs 0.007 F1, decision quality is flat across chronological quartiles, and the official test queries are not measurably farther from the training manifold than training queries are from each other, in two independent embedding spaces.
  Decomposing the misses instead splits them exactly evenly between candidates never retrieved and candidates retrieved but ranked below the cut.
  Measuring the remedies for each half, BM25 length-normalisation tuning, an event-triple view, and full-content dense fusion lift top-200 recall by three to seven points, and citation-graph features add 0.014 F1 over eight seeds once own-citation leakage is removed, while per-query cutoff rules, a zero-shot reranker swap, and a date filter do not help.
  We also document four evaluation artifacts, each of which reversed a result once the protocol was corrected.
\end{abstract}




\maketitle


\section{Introduction}

The great advance of the technology of artificial intelligence and the increasing amount of legal data have driven the improvement for diverse tasks in legal domains. One of the popular tasks is case retrieval, which searches for the relevant cases for a given query document, so that legal professionals can use and cite the retrieved cases in their final decision.

In line with this, the Competition on Legal Information Extraction and Entailment (COLIEE) 2026 Task 1 has been introduced, continuing the series described in the 2023 and 2025 overviews~\cite{coliee2023,coliee2025}.\footnote{The COLIEE 2026 overview paper had not been published at the time of writing, so we cite the most recent available overviews and report the 2026 task definition and official scores as given to participants.} The goal of the task is to find the cited cases from a set of case documents, given a query case document. In the 2026 challenge, the provided dataset consists of a corpus of 9,556 documents, where the actual cited cases have been replaced with the \fragmentsuppressedmarker{} markers. This structural constraint severely limits the effectiveness of traditional lexical matching, as the direct textual overlap between a query and its gold-standard citations is inherently diminished. Consequently, robust systems must look beyond simple keyword matching and use both semantic representations and the latent structural relations within the legal corpus.

In this paper, we propose a multi-stage hybrid pipeline to mitigate the lexical gap caused by citation suppression. Our approach integrates four distinct components: (i) a multi-view BM25 retriever that employs citation context windows to capture local relevance signals, (ii) a neural reranking stage consisting of bi-encoders, BGE-M3, and cross-encoders, to capture deep semantic interactions, (iii) a graph-based reranking stage whose goal is to capture latent structural relationships between cases through shared legal entities and propagate relevance signals across the corpus graph, and (iv) a LightGBM meta-learner that aggregates 34 features extracted from three stages to decide whether each case should be predicted as a citation or not.

Our empirical evaluation shows a notable performance discrepancy: the cross-validated F1 score on the provided training data is 0.311 while the one on the official test set is 0.177, which ranked 35th among 54 official runs (15th of the 22 participating teams by each team's best run). We conducted an ablation study for the proposed model, demonstrating that the cross-encoder is the largest single contributor to performance (24\% relative). The source code and submission materials are available at \url{https://github.com/rabqatab/coliee2026_ABAI}.

Our workshop paper attributed that gap to three causes: a first-stage recall ceiling, temporal distribution shift, and a decision threshold miscalibrated by training on gold-injected candidate pools.
It tested none of them.
This extended version does.
Under a chronological three-way split in which every decision parameter is fit off the evaluation slice, threshold transfer costs 0.007 F1, decision quality is flat across four chronological quartiles, and the official test queries turn out to be no farther from the training manifold than training queries are from each other.
None of the three causes reproduces on any internal proxy we could build.

We then ask where the loss actually is.
The missed citations divide exactly in half between candidates that never entered the candidate pool and candidates that entered it and were ranked below the cut, which makes retrieval and reranking equally worth fixing rather than retrieval alone.
We measure the remedies proposed for each half and report which of them work: tuning BM25 length normalisation, adding an event-triple retrieval view, and embedding full documents instead of truncated ones each lift top-200 recall by three to seven points, and citation-graph features add 0.014 F1 across eight seeds once a leakage path is removed, while per-query cutoff rules, an off-the-shelf reranker swap, and a date filter do not help.
Along the way four separate results reversed once we corrected the protocol that produced them, three of them in the direction we expected, and we report those as well (Section~\ref{sec:posthoc_artifacts}).

\section{Related Work}

The increasing popularity of deep learning models has encouraged the research community to develop diverse models for case retrieval tasks~\cite{feng2024survey}. A large portion of this research has focused on the structural characteristics of legal documents, identifying and exploiting the key parts for better retrieval. For example, Tran et al.~\cite{tran2019} demonstrated that extracting lexical features from different parts of query and candidate documents improved retrieval over whole-document comparison. SAILER~\cite{sailer} suggested an asymmetric encoder-decoder model that encodes the fact section and reconstructs aggressively-masked reasoning and decision sections, forcing the encoder to capture cross-section dependencies. Similarly, Ma et al.~\cite{slr} proposed Structured Legal Retrieval (SLR) framework, which first splits a legal document into functional segments (facts, reasoning, ruling), and then computes embeddings of each segment text. By incorporating an external charge-wise relation graph, which captures co-occurrence and hierarchical relationships between legal charges, the framework finally estimates the rank of a given document.

Recently, integration of graph-based models to prior work has been suggested. GraphRAG~\cite{graphrag_survey}, which models and uses the relations among legal documents, has gained significant attention. Louis et al.~\cite{louis2023} apply GNNs to statutory article retrieval, showing that propagating information through a citation graph improves dense retrieval by capturing inter-article dependencies. Zhang et al.~\cite{cfgl} propose CFGL-LCR, a counterfactual graph learning framework for legal case retrieval that disentangles causal and shortcut features on case graphs.

Our framework differs from these prior studies in several important ways. First, instead of segmenting legal documents by legal function (e.g., facts, reasoning, and ruling), the proposed framework employs citation context windows around \fragmentsuppressedmarker{} markers to form a retrieval-focused structural view. We also propose a GraphRAG Lite module that consists of two lightweight components: (i) regex-based entity extraction (statutes, judges, legal domains, outcomes) and (ii) Leiden community detection~\cite{leiden}, which together capture latent structural relations without LLM overhead or explicit citation graphs. We use Graph Attention Network~\cite{gat} to generate graph-level features.

Section~\ref{sec:posthoc} draws on three further lines, which we test rather than extend.
Ma et al.~\cite{lecut} argue that a learned, query-adaptive truncation of the ranked list beats a fixed cutoff in legal search; we implement six rules in that family and find that all of them lose to an honestly calibrated global threshold on our data.
Adaptive re-ranking over a corpus graph~\cite{macavaney2022adaptive} expands a candidate pool with graph neighbours of high-scoring seeds, which raises our recall once the budget exceeds the seed pool.
Event-based retrieval~\cite{joshi2023u} treats extracted subject-verb-object triples as a non-redundant lexical signal, and it reproduces on our corpus.

Our distribution probe follows a separate line concerned with diagnosing retrieval robustness rather than improving it.
Ko et al.~\cite{ko2025when} detect out-of-distribution \emph{corpora} for dense retrievers without relevance labels, in order to decide when a retriever needs updating.
We borrow the label-free framing and apply it on the query side instead.
That is exactly our constraint, since the official test labels were never released, and we adapt the idea into the embedding-neighbourhood probe of Section~\ref{sec:posthoc_retest}.

\section{Method}

\begin{figure}[t]
  \centering
  \includegraphics[width=\columnwidth]{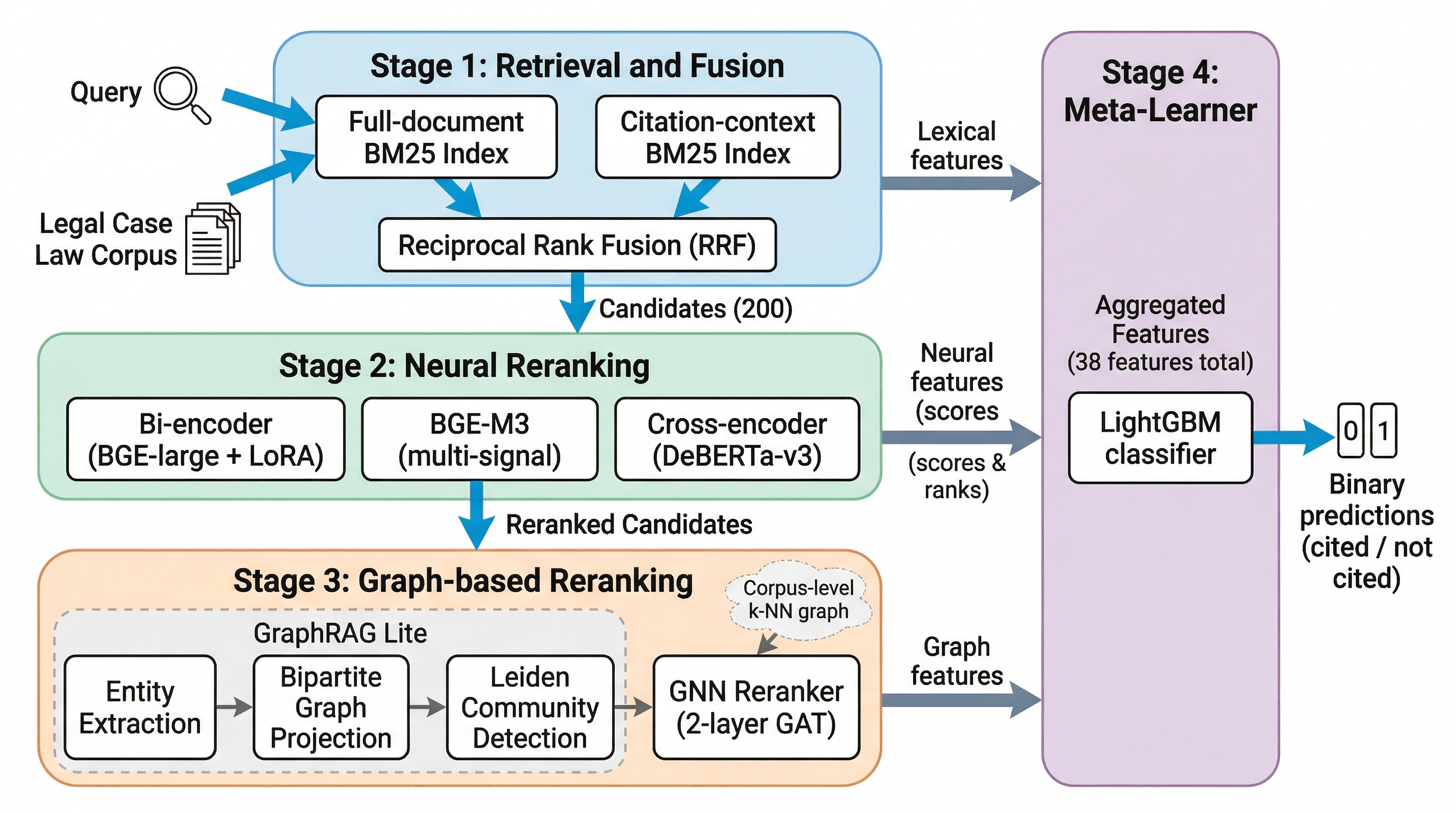}
  \caption{Architecture of the proposed multi-stage pipeline.}
  \label{fig:pipeline}
\end{figure}

Figure~\ref{fig:pipeline} illustrates the overall architecture of the proposed framework, which consists of four stages. Stages 1 to 3 both narrow the candidate set and extract features from the given query and candidate cases, which are then fed into Stage 4 for the final decision. Note that the components in each stage are trained independently on different tasks.

\subsection{Stage 1: Retrieval and Fusion}

Figure~\ref{fig:context_example} shows what a suppressed citation actually looks like in the corpus.
The cited case name is gone, but the neutral citation and the reporter reference beside it survive, as does the legal proposition the citation was offered for.
This is the asymmetry the rest of the pipeline is built around.
A retriever cannot match on the case name, and the surviving numerals and the surrounding proposition carry whatever signal remains, which is why we index a window around the marker rather than the document alone.

\begin{figure}[t]
  \small
  \setlength{\fboxsep}{6pt}
  \noindent\fbox{\parbox{0.94\columnwidth}{%
    \ldots{}The first issue is one of procedural fairness and must be reviewed according to the standard of correctness (\fragmentsuppressedmarker{}; 344 N.R. 257; 2005 FCA 404). [11] The second issue involves establishing the content and interpretation of Cameroonian law\ldots{}%
  }}
  \caption{A citation context from case 000002 of the 2026 training corpus, shown verbatim. Suppression removes the cited case name but leaves the neutral citation (\texttt{2005 FCA 404}), the reporter reference (\texttt{344 N.R. 257}), and the proposition the case was cited for. Text overlap with the cited document is therefore reduced but not eliminated, and what survives is lexical and structural rather than narrative. The $\pm$150-word window we index is centred on this marker.}
  \label{fig:context_example}
\end{figure}

Inspired by the observation that legal documents exhibit strong lexical characteristics dominated by legal terminology~\cite{feng2024survey}, we first design a lexicon-based case retrieval mechanism as the initial step for filtering case candidates. In particular, we employ BM25, which computes the similarity index between a pair of documents. Here, we calculate two indices: one for a pair of a query and a full case document, and the other for a pair of a query and a key part of a case document. We consider $\pm$150 words around a \fragmentsuppressedmarker{} marker as the key part. The rationale behind this choice is that the marker indicates the location where cited text was redacted, so the surrounding context contains the strongest relevance signal for identifying the cited case. Note that this approach extends the strategy proposed by Tran et al.~\cite{tran2019}, who suggested extracting lexical features from different structural parts of legal documents for improved retrieval, rather than comparing whole documents. Throughout the process, we select top 200 and 30 documents in terms of BM25 indices for full document and key part. Both indices are then fused by Reciprocal Rank Fusion~\cite{rrf} (RRF), which is robust to score distribution mismatches between heterogeneous retrieval systems. Formally, for a candidate document $d$ ranked at position $r_i(d)$ under index $i \in \{\text{full}, \text{ctx}\}$, the RRF score is
\[
  \text{RRF}(d) = \sum_{i \in \{\text{full}, \text{ctx}\}} \frac{1}{k + r_i(d)},
\]
with $k=60$. Top 200 candidate cases are finally selected and forwarded to Stage 2.

\noindent\textbf{Computed Features.}
In addition to BM25 scores, we compute five lexical feature values (cosine similarity of TF-IDF vectors, Jaccard coefficient, shared bigrams, length ratio, and shared legal terms) for all query--document pairs based on token-level statistics.
We also extract two citation-context features per candidate: the number of citation context windows whose BM25 top-30 results include that candidate, and the maximum BM25 score across those windows.
Together, Stage~1 contributes 9~features to the meta-learner (Table~\ref{tab:features}, \#1--9).

\subsection{Stage 2: Neural Reranking}
\label{sec:neural}

Stage 2 is responsible for reranking the candidate cases and extracting neural features. To this end, we design three components: a bi-encoder, BGE-M3, and a cross-encoder.

\noindent\textbf{Bi-encoder.}
We fine-tune BGE-large-en-v1.5~\cite{bge} with LoRA~\cite{hulora} ($r$=16, $\alpha$=32) as a \emph{contrastive retrieval model} (trained independently from other stages). Given a query $q$, a positive document $d^+$ (a gold citation), and 7 hard negatives $\{d^-_1, \ldots, d^-_7\}$ sampled from the BM25 top-50 non-gold candidates, the per-query InfoNCE loss is
\[
  \mathcal{L}_\text{NCE}(q) = -\log \frac{\exp(s\,\cos(e_q, e_{d^+}))}{\exp(s\,\cos(e_q, e_{d^+})) + \sum_{i=1}^{7} \exp(s\,\cos(e_q, e_{d^-_i}))},
\]
where $e_q$ and $e_d$ are the query and document embeddings produced by the LoRA-adapted encoder and $s$ is the scale factor. The loss pulls the cosine similarity $\cos(e_q, e_{d^+})$ above that of every hard negative. Training uses AdamW for 3 epochs (lr=2e-5, batch size 16).

\noindent\textbf{BGE-M3.}
To capture complementary retrieval signals beyond dense cosine similarity, we additionally score candidates using BAAI/bge-m3~\cite{bgem3} \emph{without further fine-tuning}; the released checkpoint provides dense, sparse, and ColBERT similarity signals within a single model. The sparse signal recovers exact lexical matches that dense retrieval may miss, while ColBERT captures fine-grained token-level interactions without full cross-attention cost.

\noindent\textbf{Cross-encoder.}
We select DeBERTa-v3-large~\cite{deberta} as our cross-encoder because its disentangled attention mechanism explicitly models content and positional information separately, which is beneficial for matching legal facts across structurally different case documents. It is fine-tuned as a binary classifier with cross-entropy loss using AdamW for 3 epochs (lr=1e-5, batch size 16). Because legal documents (2,000--5,000 words) far exceed the 512-token input limit, naive truncation discards most of the document. We therefore use \emph{selective truncation}: for each query-candidate pair, we concatenate document head/tail (100 words each) with the top-5 citation context windows and top-10 paragraphs by TF-IDF similarity, within a 500-word budget. This prioritizes the passages most likely to contain citation-relevant information over arbitrary positional truncation, following the intuition from context-aware passage selection~\cite{dai2020context} and structured legal
retrieval~\cite{slr}.

\noindent\textbf{Computed Features.}
Each neural model produces a relevance \emph{score} and a within-query \emph{rank} (the ordinal position when all candidates for the same query are sorted by that model's score). The rank is a derived feature computed after scoring, not a direct model output. Stage~2 contributes 8~features in total: bi-encoder score and rank, four BGE-M3 signal scores, and cross-encoder score and rank (Table~\ref{tab:features}, \#10--17).

\subsection{Stage 3: Graph-Based Reranking}

While citation graphs are the natural structure for case law retrieval~\cite{lesicin}, the citation information is suppressed in our setting, and explicit citation links therefore cannot be used. We instead construct proxy citation structures through shared legal entities, following the GraphRAG paradigm~\cite{graphrag_survey}. Stage~3 consists of two independent sub-modules, GraphRAG Lite and a GNN reranker, that operate on \emph{different graphs} and produce complementary features.

\subsubsection{GraphRAG Lite.}\hfill\break

\noindent\textbf{Entity extraction.}
We extract four entity types from each document via regex: statute citations (e.g., ``R.S.C.\ 1985, c.~I-2, s.~3(1)''), judge names (e.g., ``Gagn\'{e} J.''), legal domain (8 categories by keyword matching: immigration, tax, IP, aboriginal, criminal, administrative, labour, environmental), and case outcome (dismissed/allowed).

\noindent\textbf{Bipartite graph.}
Let $D = \{d_1, \ldots, d_n\}$ ($n{=}9{,}556$) denote documents and $S = \{s_1, \ldots, s_m\}$ the extracted entities. We construct a weighted bipartite graph $G_{\text{bip}} = (D \cup S, E)$ where an edge $(d, s) \in E$ exists if document $d$ contains entity $s$. Each edge is weighted by entity type:
\[
  w(d, s) = w_{\tau(s)} \cdot \mathbf{1}[(d, s) \in E]
\]
with type weights $w_\text{statute}=0.50$, $w_\text{judge}=0.15$, $w_\text{domain}=0.10$, and $w_\text{outcome}=0.05$.

\noindent\textbf{Document graph projection.}
We project $G_{\text{bip}}$ onto a document--document graph $G_{\text{doc}} = (D, E_{\text{doc}})$ via weighted one-mode projection:
\[
  A_{\text{doc}}(d_i, d_j) = \sum_{s \in S} w(d_i, s) \cdot w(d_j, s)
\]
Two documents sharing more (or higher-weighted) entities receive a stronger edge.

\noindent\textbf{Community detection.}
Leiden community detection~\cite{leiden} is applied to $G_{\text{doc}}$ at three resolutions ($\gamma = 0.5, 1.0, 2.0$), each producing a complete partition of all 9,556 documents into non-overlapping communities. Lower resolutions yield larger communities (broad legal domains); higher resolutions yield smaller clusters (fine-grained case groups). We use all three partitions.

\noindent\textbf{Computed Features.}
For each query--candidate pair $(q, c)$, GraphRAG Lite produces 9 features: three same-community indicators (one per resolution, with value 1 if $q$ and $c$ belong to the same community and 0 otherwise), the community Jaccard (the fraction of the three resolutions where $q$ and $c$ co-occur), the shared statute count and shared judge count, the same-domain and same-outcome indicators, and an entity overlap score computed as a weighted Jaccard over all shared entities (Table~\ref{tab:features}, \#18--26).

\subsubsection{GNN Reranker.}\hfill\break

The GNN reranker operates on a corpus-level graph $G_{\text{knn}}$ that is distinct from the entity-projected graph $G_{\text{doc}}$ used for community detection.

\noindent\textbf{Graph construction.}
$G_{\text{knn}}$ 
has 9,556 document nodes and 61,645 edges constructed by combining two edge sources:
(1)~$k$-nearest-neighbor edges ($k{=}8$) based on cosine similarity of bi-encoder embeddings, capturing semantic relatedness; and
(2)~entity overlap edges from $G_{\text{doc}}$, weighted at 30\% of total edge weight.

\noindent\textbf{Node features.}
Each node is represented by a 35-dimensional feature vector: 32 dimensions from PCA-compressed bi-encoder embeddings concatenated with 3 retrieval scores (BM25 RRF, bi-encoder similarity, cross-encoder probability).

\noindent\textbf{Architecture.}
A 2-layer Graph Attention Network~\cite{gat} ($d_\text{hidden}{=}64$, 4 attention heads, dropout 0.1) processes $G_{\text{knn}}$. We choose GAT over GCN because its attention mechanism learns to weight neighbor contributions differently, distinguishing highly relevant corpus neighbors from coincidentally similar documents~\cite{louis2023}. The GNN outputs a relevance score per document node, trained with binary cross-entropy loss on gold citation pairs using Adam for 50 epochs (lr=1e-3, full-batch transductive). To avoid information leakage, we use 5-fold out-of-fold predictions.

\noindent\textbf{Computed Features.}
The GNN produces 2 features per query--candidate pair: the GNN relevance score and the within-query GNN rank. In total, Stage~3 contributes 11 features to the meta-learner (Table~\ref{tab:features}, \#18--28).


\subsection{Stage 4: Meta-Learner}
\label{sec:meta_learner}

We adopt a learning-to-rank meta-learner rather than a fixed score combination because the relative importance of retrieval signals varies across legal domains: statute overlap matters more for tax cases, while judge co-citation matters more for immigration cases~\cite{feng2024survey}. A gradient-boosted tree captures these non-linear feature interactions without manual weight tuning. A LightGBM~\cite{lgbm} binary classifier trained with binary cross-entropy loss, chosen for its ability to handle heterogeneous feature types (continuous scores, binary indicators, counts) and its robustness to the class imbalance inherent in retrieval~\cite{lgbm} (typically 4 positives among 200 candidates per query). It combines 34 features from all preceding stages (Table~\ref{tab:features}).

Table~\ref{tab:features} lists all 34 features grouped by pipeline stage.

\begin{table*}[t]
  \caption{Feature inventory of the LightGBM meta-learner (34 features per query--candidate pair). Type: cont.\ = continuous, int = integer count or rank, bin.\ = binary.}
  \label{tab:features}
  \begin{threeparttable}
    \small
    \begin{tabular}{@{}clllp{5.8cm}@{}}
      \toprule
      \textbf{Stage} & \textbf{\#} & \textbf{Feature} & \textbf{Type} & \textbf{Description} \\
      \midrule
      \multirow{9}{*}{\rotatebox{90}{\scriptsize\textbf{Stage 1}}}
      & 1 & bm25\_score & cont. & Raw BM25 score (full-document query) \\
      & 2 & bm25\_rrf\_score & cont. & RRF-fused score from multi-view BM25 \\
      & 3 & n\_context\_matches & int & Number of citation-context windows matching $c$ \\
      & 4 & max\_context\_bm25 & cont. & Highest BM25 score among context windows for $c$ \\
      & 5 & tfidf\_cosine & cont. & Cosine similarity of TF-IDF vectors (50K vocab) \\
      & 6 & jaccard & cont. & Word-level Jaccard: $|T_q \cap T_c| / |T_q \cup T_c|$ \\
      & 7 & shared\_bigrams & cont. & Bigram-level Jaccard overlap \\
      & 8 & length\_ratio & cont. & $\min(|q|,|c|) / \max(|q|,|c|)$ in words \\
      & 9 & shared\_legal\_terms & int & Shared words from 42-term legal vocabulary \\
      \midrule
      \multirow{8}{*}{\rotatebox{90}{\scriptsize\textbf{Stage 2}}}
      & 10 & biencoder\_score & cont. & Cosine similarity of BGE-large embeddings \\
      & 11 & biencoder\_rank & int & Within-query rank by bi-encoder score \\
      & 12 & m3\_dense\_score & cont. & BGE-M3 dense embedding similarity \\
      & 13 & m3\_sparse\_score & cont. & BGE-M3 sparse lexical matching score \\
      & 14 & m3\_colbert\_score & cont. & BGE-M3 ColBERT late-interaction score \\
      & 15 & m3\_fused\_score & cont. & Weighted fusion of M3 dense/sparse/ColBERT \\
      & 16 & crossencoder\_score & cont. & DeBERTa softmax $P(\text{cited} \mid q, c)$ \\
      & 17 & crossencoder\_rank & int & Within-query rank by cross-encoder score \\
      \midrule
      \multirow{11}{*}{\rotatebox{90}{\scriptsize\textbf{Stage 3}}}
      & 18 & same\_community\_{0.5} & bin. & Co-membership at Leiden resolution 0.5 \\
      & 19 & same\_community\_{1.0} & bin. & Co-membership at Leiden resolution 1.0 \\
      & 20 & same\_community\_{2.0} & bin. & Co-membership at Leiden resolution 2.0 \\
      & 21 & community\_jaccard & cont. & Fraction of resolutions with co-membership \\
      & 22 & shared\_statutes & int & Count of shared statute citations \\
      & 23 & shared\_judges & int & Count of shared judge names \\
      & 24 & same\_domain & bin. & Same legal domain (8 categories) \\
      & 25 & same\_outcome & bin. & Same case outcome (dismissed/allowed) \\
      & 26 & entity\_overlap\_score & cont. & Weighted Jaccard over all shared entities \\
      & 27 & gnn\_score & cont. & GAT output relevance score \\
      & 28 & gnn\_rank & int & Within-query rank by GNN score \\
      \midrule
      \multirow{10}{*}{\rotatebox{90}{\scriptsize\textbf{Stage 4}}}
      & 29 & bm25\_rrf\_rank\_norm & cont. & Normalized BM25 rank within query (0=best) \\
      & 30 & bm25\_rrf\_score\_gap & cont. & BM25 score minus query mean \\
      & 31 & biencoder\_score\_gap & cont. & Bi-encoder score minus query mean \\
      & 32 & crossencoder\_score\_gap & cont. & Cross-encoder score minus query mean \\
      & 33 & top\_score\_ratio & cont. & BM25 score / query maximum score \\
      & 34 & score\_above\_median & bin. & BM25 score $>$ query median \\
      \bottomrule
    \end{tabular}
  \end{threeparttable}
\end{table*}

Training uses GroupKFold (5 folds, grouped by query), with 3 random seeds for stability (15 models total).
Key training strategies include:
(1) gold positive injection, which adds the 3,484 gold positives lying outside the BM25 top-200 to the candidate pool;
(2) stratified negative sampling (50\% hard / 30\% medium / 20\% easy, 10:1 ratio, cap 50 per query);
and (3) threshold re-optimization on the full candidate pool.
Letting $\phi(q, c) \in \mathbb{R}^{34}$ denote the feature vector for the query--candidate pair $(q, c)$ and $f$ the trained LightGBM probability, the predicted citation set for query $q$ is
\[
  \hat{C}_q = \{\, c \in C_q : f(\phi(q, c)) \geq t \,\},
\]
where $C_q$ is the BM25-RRF top-200 candidate pool from Stage~1 and the threshold $t$ is optimised to maximise micro-F1 on training data.

\noindent\textbf{Scale and regularisation.}
Although the dataset contains 2{,}001 training queries, the meta-learner is trained at the granularity of query--candidate pairs: 2{,}001 queries with up to 200 candidates per query yield on the order of $4{\times}10^{5}$ pair-level examples after the stratified negative-sampling cap.
The ratio of training examples to features (roughly $1.2{\times}10^{4}$ examples per feature) places the meta-learner in the regime where gradient-boosted trees are robust to feature dimensionality.
We further mitigate overfitting through query-grouped cross-validation (GroupKFold by query ID prevents within-query leakage across folds), early stopping at 80 rounds on a per-fold validation set, feature and row subsampling (column and row sampling rates both set to 0.8, minimum samples per leaf set to 20), and averaging across three random seeds.

\noindent\textbf{Hyperparameter choices.}
Several hyperparameters were fixed without ablation due to compute budget. Their values are motivated as follows.
The citation-context window of $\pm 150$ words follows Tran et al.~\cite{tran2019} and corresponds to a typical paragraph length in Canadian federal case judgments.
The entity type weights $w_\text{statute}=0.50$, $w_\text{judge}=0.15$, $w_\text{domain}=0.10$, $w_\text{outcome}=0.05$ reflect an information-content ordering: statute citations are the most discriminative legal entities (typically unique to a doctrinal area), whereas case outcomes (dismissed/allowed) are coarse binary attributes that appear in roughly half of all judgments and are the least discriminative.
The Leiden resolutions $\gamma \in \{0.5, 1.0, 2.0\}$ form a $\sqrt{2}$-spaced sweep chosen to capture broad legal domains (low $\gamma$) and fine-grained case clusters (high $\gamma$) without manual tuning; using all three simultaneously is robust to misspecification of any single resolution.
The 7 hard negatives per positive in bi-encoder training follow the default of the BGE recipe~\cite{bge}.
A systematic sensitivity sweep of these hyperparameters is left for future work.

\section{Results}

\subsection{Internal Validation}

Table~\ref{tab:internal} compares our pipeline against simplified reproductions of prior COLIEE winning systems on the same candidate pool and evaluation protocol, isolating the contribution of each reranking strategy.
We do not retrain the original systems end-to-end; instead, we reimplement the published feature set and reranking pipeline of each baseline as follows.
\textbf{BM25 (vanilla)} is Okapi BM25 over full documents only, without RRF, without neural reranking, returning the top-200 candidates per query and predicting by a single global score threshold tuned on the validation set.
\textbf{TQM 2024 (LTR)} combines BM25 with a LightGBM ranker over the 14-feature subset reported in the TQM system paper~\cite{li2024towards} (BM25, TF-IDF cosine, document and query lengths, simple lexical overlaps, and a single SBERT cosine), without SAILER, graph features, or our citation-context-aware cross-encoder.
\textbf{JNLP 2025 (BM25+SAILER)} uses BM25 as the first stage and SAILER~\cite{sailer} as a reranker (using the publicly released checkpoint without retraining), followed by a LightGBM aggregator over the 8-feature set described by the JNLP authors~\cite{jnlp2025}.
The three baselines use the same training/validation split, the same candidate pool size (top-200 from BM25), and the same LightGBM hyperparameters where applicable, so differences among them in Table~\ref{tab:internal} reflect the published feature differences rather than tuning. The row for our system in the same table is included for ease of reference and follows a different protocol described below.

We use two complementary evaluation protocols.
For the baseline comparisons reported in Table~\ref{tab:internal}, we hold out the most recent 20\% of training queries (401 of 2{,}001) by case ID as a chronological validation set, training each system on the remaining 1{,}600 queries.
For the ablation and feature analyses reported in Tables~\ref{tab:ablation} and following, we use 5-fold GroupKFold cross-validation on all 2{,}001 training queries, grouped by query ID to prevent within-query leakage, and report out-of-fold (OOF) micro-F1.
The F1=0.311 reported for our system in Table~\ref{tab:internal} is the 5-fold OOF result on the full 2{,}001-query training set, reproduced in the table for direct comparison; a chronological hold-out evaluation that retrains the meta-learner on the earliest 80\% of queries and evaluates on the most recent 20\% would more directly quantify temporal generalisation, and we discuss this limitation in Section~\ref{sec:cv_test_gap}.

We submitted three runs at different thresholds: \emph{balanced} ($t$=0.547, the CV-optimal threshold), \emph{recall} ($t$=0.400, favoring coverage), and \emph{precision} ($t$=0.700, favoring accuracy).

\begin{table}[h]
  \caption{Internal validation results against simplified prior winners.}
  \label{tab:internal}
  \begin{tabular}{lccc}
    \toprule
    System & F1 & Prec. & Recall \\
    \midrule
    BM25 (vanilla) & 0.033 & 0.018 & 0.217 \\
    TQM 2024 (LTR) & 0.119 & 0.145 & 0.101 \\
    JNLP 2025 (BM25+SAILER) & 0.138 & 0.123 & 0.157 \\
    \midrule
    Ours ($t$=0.547, balanced) & 0.311 & 0.368 & 0.270 \\
    Ours ($t$=0.400, recall) & 0.285 & 0.264 & 0.310 \\
    Ours ($t$=0.700, precision) & 0.273 & 0.472 & 0.192 \\
    \bottomrule
  \end{tabular}
\end{table}

\subsection{Official Test Results}
\label{sec:official_test}

Table~\ref{tab:official} shows our three submissions alongside the top performers among the 22 participating teams, which together produced 54 official runs.
Our best run (run~2, recall-favouring) reached F1=0.177 and ranked 35th of the 54 per-run submissions (15th among the 22 teams when each team is represented by its best run), substantially below our cross-validated estimate of F1=0.311.
The winner (NOWJ) achieved F1=0.422, and JNLP, the COLIEE 2025 winner at F1=0.335, improved to 0.413.

\begin{table}[h]
  \caption{Official COLIEE 2026 Task 1 results (selected runs).}
  \label{tab:official}
  \begin{tabular}{llccc}
    \toprule
    Rank & Team / Run & F1 & Prec. & Recall \\
    \midrule
    1 & NOWJ / sub\_2 & 0.422 & 0.424 & 0.421 \\
    3 & JNLP / random\_forest & 0.413 & 0.434 & 0.393 \\
    7 & SIL / sil2 & 0.387 & 0.419 & 0.360 \\
    \midrule
    35 & ABAI / run2 (recall) & 0.177 & 0.160 & 0.199 \\
    36 & ABAI / run1 (balanced) & 0.167 & 0.195 & 0.146 \\
    41 & ABAI / run3 (precision) & 0.131 & 0.220 & 0.093 \\
    \bottomrule
  \end{tabular}
\end{table}

\subsection{Per-Stage Retrieval Quality}
\label{sec:stage_quality}

A natural diagnostic for a cascaded retrieval pipeline is how recall and ranking quality evolve across stages.
By construction, every reranker and feature extractor in Stages~2--3 operates on the fixed pool of 200 candidates produced by Stage~1, so the downstream stages can only change the ordering of candidates within the pool, not the recall of the pool itself.
Recall@200 is therefore determined entirely by the BM25-RRF first stage, which retrieves 57.8\% of gold positives on the training data; this value is an upper bound on the achievable recall of every subsequent stage in our system.
The reranking contribution of each stage is captured by the cross-validated F1 deltas reported in the ablation (Table~\ref{tab:ablation}), which measure the impact of each component on the final binary decision.
Test-set recall@$k$ and NDCG@$k$ per stage would have been the most informative diagnostic for first-stage failures; the gold labels for the official test set were not released to participants in time for this analysis, and we report only the official test F1 in Section~\ref{sec:official_test}.

\subsection{Ablation Study}

Table~\ref{tab:ablation} shows the contribution of each pipeline component, measured by cross-validation with out-of-fold predictions on RRF-only candidate pools.

\begin{table}[h]
  \caption{Component ablation (5-fold cross-validation). Feature \# refers to Table~\ref{tab:features}.}
  \label{tab:ablation}
  \begin{tabular}{llccr}
    \toprule
    Configuration & Features & F1 & Recall & $\Delta$F1 \\
    \midrule
    Full pipeline & all 34 & 0.311 & 0.270 & -- \\
    \midrule
    $-$ Cross-encoder & \#16, 17 & 0.236 & 0.248 & $-$0.075 \\
    $-$ Lexical features & \#5--9 & 0.289 & 0.276 & $-$0.023 \\
    $-$ GNN + BGE-M3 & \#12--15, 27, 28 & 0.299 & 0.286 & $-$0.012 \\
    $-$ GraphRAG & \#18--26 & 0.305 & 0.295 & $-$0.006 \\
    $-$ Bi-encoder & \#10, 11 & 0.310 & 0.281 & $-$0.001 \\
    \bottomrule
  \end{tabular}
\end{table}

The cross-encoder is the dominant component ($-$24\% relative F1).
Lexical features rank second ($-$7.3\%), followed by GNN + BGE-M3 ($-$3.8\%) and GraphRAG ($-$2.0\%).
The bi-encoder contributes minimally ($-$0.3\%), as its signal is largely subsumed by the cross-encoder and BM25.

The GraphRAG Lite and GNN reranker components together contribute $\Delta$F1$\,\approx\,-0.012$ on cross-validation.
Their training-time cost is modest: entity extraction is regex-only, Leiden community detection runs once on the 9{,}556-document graph in roughly 107 seconds on a single CPU core, and the GNN trains in approximately 25 minutes on a single NVIDIA DGX Spark GB10 GPU (one-off; the model is cached for inference).
Inference cost on the same hardware is negligible: community lookups are $O(1)$ per pair and a single forward pass of the 9{,}556-node GAT completes in about 3 seconds for the 400 test queries.
The graph components are therefore best characterised as inexpensive structural retrofits to the meta-learner feature set, rather than as core ranking components.

To quantify the discriminative signal carried by graph features, we compare their distribution across gold and non-gold candidate pairs within the BM25 top-200 candidate pool on the training data (Table~\ref{tab:graph_enrichment}).
Graph features show consistent enrichment in gold pairs across all four categories, with ratios of 1.43--1.72$\times$ over non-gold pairs.
This signal is real but modest, and it is correlated with rather than redundant to lexical similarity, consistent with the additive contribution of graph features in the ablation.

\begin{table}[h]
  \caption{GraphRAG feature distribution in gold vs.\ non-gold pairs (BM25 top-200, training data). Feature \# refers to Table~\ref{tab:features}.}
  \label{tab:graph_enrichment}
  \begin{tabular}{clccc}
    \toprule
    \# & Feature & Gold & Non-gold & Ratio \\
    \midrule
    20 & same\_community ($\gamma=2.0$) & 26.8\% & 17.3\% & 1.55$\times$ \\
    23 & shared\_judges (mean count) & 0.22 & 0.13 & 1.72$\times$ \\
    22 & shared\_statutes (mean count) & 0.90 & 0.59 & 1.53$\times$ \\
    26 & entity\_overlap (mean score) & 0.171 & 0.120 & 1.43$\times$ \\
    \bottomrule
  \end{tabular}
\end{table}

The niche in which graph features help most is the deeper portion of the BM25-RRF candidate pool: 470 gold pairs that BM25-RRF ranks below position 50 still exhibit a strongly positive aggregate graph signal (same community at all three resolutions, multiple shared statutes, and shared judges).
A representative example is the citation from query 057595 (\emph{Olvera v.\ Canada (M.C.I.)}, 2012, a Federal Court judicial review of a Refugee Protection Division decision concerning a Mexican family) to gold case 033043 (a 2010 Federal Court refugee-protection decision in the same area decided by the same presiding judge). BM25-RRF ranks 033043 only 63rd for this query, yet GraphRAG features capture the latent affinity through three shared Leiden communities, three shared statute citations, two shared judge names, and the same legal domain.
Such pairs are precisely the cases that pure semantic reranking is least able to recover, and graph features provide an inexpensive complementary signal for the meta-learner.

\subsection{Analysis: CV-to-Test Gap}
\label{sec:cv_test_gap}

The 43\% relative drop from CV F1=0.311 to test F1=0.177 is the primary finding of our work. We identify three candidate factors and report the empirical basis available to us at submission time. We state them here as the hypotheses we formed then, not as established causes: Section~\ref{sec:posthoc} re-tests all three under leakage-free protocols, and none of them reproduces.

\textbf{Recall ceiling.}
On the training queries, only 57.8\% (4{,}767 of 8{,}251) of gold positives appear in the BM25 top-200 candidate pool, so the achievable post-reranking recall is bounded above by this fraction even for a perfect downstream reranker.
This ceiling is a property of the first-stage retrieval design rather than of any particular evaluation split: our cross-validated post-threshold recall of 0.270 already sits well below the ceiling, and the test recall of 0.199 (run~2, the recall-favouring run) leaves further headroom against the same upper bound.
The recall ceiling therefore acts as a baseline constraint shared by both cross-validation and test, and the drop from 0.270 to 0.199 in observed recall is attributable not to a collapse of the ceiling but to the two factors discussed below.

\textbf{Temporal distribution shift.}
Citation patterns in Canadian federal case law evolve as new statutes take effect and doctrinal interpretations are refined.
The 2026 test queries are drawn from the most recent cases in the corpus, while our training queries span a broader and earlier time range.
We partition the 2{,}001 training queries into four equal quartiles by case ID (a proxy for chronological order) and find that the BM25-RRF top-200 recall is reasonably stable across the partition: 53.7\%, 57.6\%, 59.7\%, and 60.2\% from oldest to newest quartile, with average gold citations per query in a narrow band of 3.89--4.28.
First-stage retrieval quality therefore does not appear to degrade systematically with case age within the training window, suggesting the temporal component of the CV-to-test gap is driven primarily by changes in the discriminative geometry of the reranking and threshold stages rather than by a collapsing recall ceiling.
Our 5-fold cross-validation mixes earlier and later cases in each fold, which can still overestimate generalisation to a strictly temporally out-of-distribution test split; a chronological hold-out evaluation that retrains the meta-learner on the earliest 80\% of queries and evaluates on the most recent 20\% would quantify this effect directly, and is left as future work.

\textbf{Threshold miscalibration under gold injection.}
During training, we injected the 3{,}484 gold positives that lie outside the BM25 top-200 into the candidate pool (Section~\ref{sec:meta_learner}) and optimised the decision threshold on this gold-injected pool.
At inference time, no such injection occurs and the LightGBM score distribution shifts downward: on the test queries the meta-learner output scores have a median of 0.021, a 95th percentile of 0.231, and a 99th percentile of 0.635.
The threshold $t=0.547$ that maximised training F1 therefore selects only the most confident predictions at inference: run~1 produced 1{,}316 predictions in total (mean 3.3 per query), against a training gold density of 4.1 per query.
This shift accounts for the lower precision and recall of the balanced run on the test set relative to its cross-validated estimate, and is consistent with the recall-favouring run~2 at $t=0.400$ being our best official submission.
Ma et al.~\cite{lecut} formalise distribution-aware list truncation for ranking, and adopting such calibration is a natural next step.

\section{Post-Hoc Analysis}
\label{sec:posthoc}

Section~\ref{sec:cv_test_gap} named three candidate causes for our 43\% drop and stopped there.
This section tests each one.
It then measures the remedies that the COLIEE field and the retrieval literature propose for each, and reports which of them move the number on our system.
Two of the causes we published do not survive that testing.
A further four results reversed during the study itself, once we corrected the protocol that had produced them.

\subsection{Protocol}
\label{sec:posthoc_protocol}

Every experiment below runs on the BM25-RRF candidate pool alone.
We do not inject the gold positives that fall outside that pool, which the meta-learner training of Section~\ref{sec:meta_learner} does.
Absolute F1 values in this section are therefore not comparable to the 0.311 of Table~\ref{tab:internal}: they measure a harder and more honest condition.

We sort the 2,001 training queries by case ID, a chronological proxy, and cut them into three slices: 1,400 core, 200 calibration, and 401 test.
The model trains on core only.
Every threshold and every decision-rule parameter is fit on calibration.
The test slice is read once, to report the number.
Rows we label \emph{oracle} fit their parameter on the test slice itself; they bound what a perfect choice would have given and are not deployable.

Meta-learner comparisons run over eight LightGBM seeds and we test the deltas with a paired Wilcoxon signed-rank test.
The first-stage recall numbers carry no error bars because nothing in that pipeline is stochastic: BM25 scoring, rank fusion, and embedding a fixed corpus are deterministic given the same inputs, so repeated runs return identical values.
Variance enters only where a model is trained, which is why the seed protocol applies to the meta-learner alone.
This matters more than it may appear.
A single configuration varies by 0.006 to 0.009 F1 across seeds, which is larger than the effects we report, so an unpaired reading would discard a real gain.
The paired per-seed difference has a standard deviation of 0.005, because bagging and feature subsampling push both configurations the same way at a fixed seed.
Eight seeds is the minimum that can reach significance at all, since the signed-rank test bottoms out at $p=0.0625$ for five.

The cost of this section is worth stating, because it bears on which remedies are worth adopting.
All meta-learner experiments run on CPU: one three-way evaluation takes about 25 seconds, so the eight-seed sweep over three configurations costs roughly ten minutes.
The BM25 parameter sweep takes about 25 minutes on CPU, event-triple extraction over all 9,556 documents takes five minutes, and the corpus-graph expansion takes five seconds on cached embeddings.
The two embedding probes run on CPU as well, the second taking about 25 minutes for 2,401 query documents through a six-layer encoder.
Only two experiments need a GPU: the alternative-reranker comparison over 100 queries takes 13 minutes, and embedding the corpus at 4,096 tokens takes about 6.7 hours on one NVIDIA DGX Spark GB10.
The single most effective recall lever we found is also the cheapest, and the total compute for this section is dominated by the one experiment we would not need to repeat.
We also note that the project as a whole consumed considerably more compute than these figures, including the failed and superseded runs described in Section~\ref{sec:posthoc_artifacts}.

One caveat applies throughout.
We evaluate many configurations against the same newest-20\% slice, which is adaptive test-set reuse, and repeated selection against one split inflates the apparent quality of whatever wins.
We report it rather than hide it, and we treat differences smaller than the seed spread as unresolved.

\subsection{Re-testing the Three Diagnosed Causes}
\label{sec:posthoc_retest}

To test temporal shift we split the queries into four chronological quartiles and held each out in turn, training on the other three and transferring the threshold without peeking.
First-stage recall rises gently across the quartiles, from 53.7\% on the oldest to 60.2\% on the newest, a spread of 6.5 points.
Decision F1 stays flat at 0.327, 0.330, 0.344, and 0.349, a spread of 0.022.
Both retrieval and decision are stable along the internal time axis.
If citation practice drifted enough to cost us 0.13 F1, that drift does not appear inside our training window.

To test threshold miscalibration we fit the threshold on the calibration slice and applied it to the test slice.
It reached F1=0.345 at $t=0.88$.
An oracle threshold fit on the test slice itself reached 0.352 at $t=0.90$.
The cost of not knowing the right threshold in advance is therefore 0.007 F1.
Our workshop paper treated threshold transfer as a principal cause of the drop, and on this evidence it is not one.

To test whether the official test queries are distributionally shifted we embedded all 2,001 training and 400 test query documents and measured each query's mean cosine to its five nearest training queries.
Training queries score 0.823 against each other under leave-one-out.
Test queries score 0.829.
Cohen's $d$ is $-0.16$, pointing the wrong way for a shift, and only 2.75\% of test queries fall below the training 5th percentile, against the 5\% a matched distribution would give.
The test queries are, if anything, marginally more central than the training queries.

Because a single embedding space cannot carry a negative result, we repeated the probe with a different model lineage and its own pooling rule.
A MiniLM sentence encoder with mean pooling puts training queries at 0.798 and test queries at 0.798, a Cohen's $d$ of $-0.004$.
Before reading that we checked the space is not degenerate: the mean pairwise cosine among training queries is 0.54 with a standard deviation of 0.14, so the embeddings are neither collapsed nor random.
The two embedders agree that the test queries are not farther from the training manifold, and they disagree about the sign of a very small residual effect: 2.75\% of test queries fall below the training 5th percentile under the first embedder and 6.25\% under the second, against the 5\% a matched distribution gives.
We therefore claim the negative result and not the direction.
Whatever separates the official test condition from our training data, it is not visible in query text.

Table~\ref{tab:retest} collects the three results.
None of the causes we proposed reproduces, and we are careful about what that licenses.
Each experiment tests an internal proxy built from data we hold, not the shift that actually produced F1=0.177, which we cannot observe because the official labels were never released.
The honest reading is that these three mechanisms do not reproduce on any proxy we could construct, which narrows the space of explanations without closing it.
Section~\ref{sec:limitations} names what remains.

\begin{table}[t]
  \caption{The three causes proposed in Section~\ref{sec:cv_test_gap}, each re-tested under a leakage-free protocol. None reproduces on an internal proxy. Effects are in the units of their own test; the query-text row gives both embedding spaces.}
  \label{tab:retest}
  \begin{tabular}{llr}
    \toprule
    Proposed cause & Internal proxy & Effect \\
    \midrule
    Temporal shift        & leave-one-quartile-out F1  & 0.022 spread \\
    Threshold transfer    & calibration vs oracle $t$  & 0.007 F1 \\
    Query-text shift      & 5-NN cosine to train       & $d=-0.16$ / $-0.00$ \\
    \bottomrule
  \end{tabular}
\end{table}

\subsection{Where the Pipeline Actually Fails}
\label{sec:posthoc_errors}

The test slice contains 1,643 gold citations.
The best model recovers 369 of them.
Of the 1,274 it misses, 637 never entered the top-200 candidate pool and 637 entered it and were ranked below the decision cut.
The split is exact, and it corrects the emphasis of our workshop analysis: we treated the recall ceiling as the dominant constraint, and it accounts for half the damage.

The second half deserves care, because it is easy to over-read.
Our cross-encoder runs in a selective mode that scores roughly 50 of the 200 pool candidates rather than all of them.
On a 100-query subset, 105 of the 305 in-pool golds fall outside that coverage and therefore carry a cross-encoder score of exactly zero.
A third of the golds that retrieval successfully surfaces are never seen by the component that does most of the ranking work.
Whether scoring them would recover them is a separate question, and one we cannot answer here.
The full-coverage run that would settle it did not leave a reproducible artifact.
We therefore report the coverage gap as a fact and the quality question as open, noting that the two are not exclusive and that we observe instances of both.

One case makes the second failure mode concrete.
For query 077181, an administrative-law case, the gold citation 034115 sits at rank 4 of the 200-candidate pool, so retrieval did its job.
The two documents share a statute and a legal domain, their TF-IDF cosine is 0.17, and BGE-M3 scores the pair at 0.79.
The cross-encoder scores it 0.045.
The meta-learner returns 0.169 against a threshold of 0.88, and the citation is lost.
We report this as an illustration rather than a measurement, since one case cannot apportion the 637.
It does show that some in-pool misses are golds the reranker saw and scored low, not golds it never saw.

\begin{figure*}[t]
  \centering
  \includegraphics[width=0.92\textwidth]{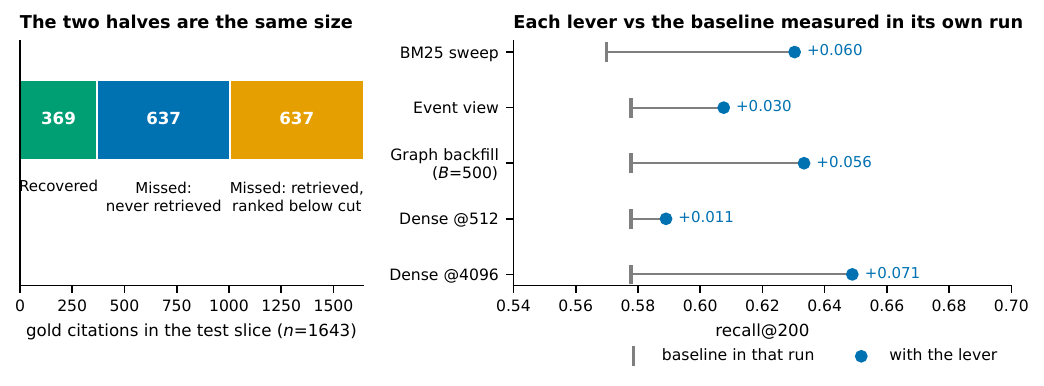}
  \caption{Left: of the 1,643 gold citations in the test slice the pipeline recovers 369, and the 1,274 it misses divide into 637 that never entered the candidate pool and 637 that entered it and were ranked below the decision cut. The two failure modes are the same size; the panel reports how large each half is, not what causes it. Right: each first-stage recall lever plotted against the baseline observed in the same run (grey tick) rather than a shared baseline, because the BM25 sweep reproduced the baseline at 0.570 while the other runs measured 0.578. Full-content dense fusion gives the largest single lift and the parameter sweep the cheapest. Generated by \texttt{scripts/experiments/make\_fig\_misses.py} from the same JSON as Table~\ref{tab:recall_levers}.}
  \label{fig:misses}
\end{figure*}

The misses also concentrate.
Administrative and immigration cases supply 769 and 639 of the 1,643 gold citations and are recovered at 21.1\% and 20.2\%, the two weakest rates among the larger domains, while tax reaches 41.0\% on 61 golds.
The composition differs too: immigration misses are ceiling-dominated at 45.9\%, whereas intellectual property misses are reranking-dominated at 52.7\%.
Which half of Figure~\ref{fig:misses} matters therefore depends on the area of law, and a single global remedy will not serve both.

The errors in the other direction are equally lopsided.
Of 1,131 predictions, 762 are false positives.
598 of those carry confidence above 0.9, and 651 sit in the pool's top five.
The meta-learner is not hedging and getting it wrong at the margin.
It is confident and wrong, on candidates that look lexically like precedents and are not, which is the failure mode a similarity-driven feature set should be expected to produce.

\subsection{Lifting the Recall Ceiling}
\label{sec:posthoc_recall}

The cheapest lever is the one we never pulled.
Sweeping the BM25 parameters raises top-200 recall from 0.570 to 0.630 in the run that measured it, and the gain comes almost entirely from the length normalisation $b$ rather than from $k_1$.
That is what a corpus of long judgments containing short relevant passages should reward: penalising length less lets a brief but decisive passage compete with a long one.
No model, no training, and no GPU is involved.

Extracting subject-verb-object triples with a dependency parser and indexing them as a further BM25 view costs five minutes of CPU time.
Fused with the existing pool it lifts recall from 0.578 to 0.608.
The 372 golds it adds across 302 queries were absent from the pool, not merely ranked low, so this is a genuine ceiling lift rather than a reordering.
The result reproduces the premise of event-based legal retrieval~\cite{joshi2023u} on a corpus it was not developed for.

Adaptive re-ranking over a corpus graph~\cite{macavaney2022adaptive} seeds a pool from the ranked list and backfills the remaining slots with graph neighbours of high-scoring seeds.
At a budget of 200 it gives us nothing, because the seed pool already fills every slot.
At 500 it reaches 0.633 and recovers 459 golds.
One implementation detail is load-bearing: graph cosine similarities fall between 0.5 and 0.9 while RRF scores fall between 0.01 and 0.05, so letting the two compete for a truncated top-$B$ evicts legitimate tail candidates and costs seven points.
The graph must backfill only.

Our first dense retrieval attempt looked like a dead end, reaching 0.435 against BM25-RRF's 0.578.
It was starved rather than unsuited.
Legal judgments run to several thousand words and their opening 380 words are largely boilerplate, so a 512-token window sees almost none of the case.
Embedding the full document instead lifts dense retrieval on its own to 0.611, and fusing it with BM25-RRF reaches 0.649, the strongest single-lever result we measured.
Lexical retrieval remains the better standalone signal, but full-content dense retrieval is complementary to it rather than redundant.

Table~\ref{tab:recall_levers} reports each lever against the baseline measured in the same run.
Four of the five share a baseline of 0.578.
The BM25 sweep is the exception: its own run reproduced the baseline at 0.570 rather than 0.578, a gap of 0.008 that its internal check accepted, and we quote its lever against its own baseline rather than against the others.
We deliberately do not report a combined figure.
The levers are plausibly complementary, since three of them recover golds the pool never held, but combining them requires re-scoring an enlarged pool and we have not done that.

\begin{table}[t]
  \caption{First-stage recall levers, each measured against the baseline observed in the same run. Lexical tuning is the cheapest gain and full-content dense fusion the largest. We do not report a combined figure because we did not measure the combination.}
  \label{tab:recall_levers}
  \begin{tabular}{lccc}
    \toprule
    Lever & Baseline & R@200 & $\Delta$ \\
    \midrule
    BM25 sweep ($k_1{=}4.5$, $b{=}1.0$)  & 0.570 & 0.630 & $+0.060$ \\
    Event-triple view, fused             & 0.578 & 0.608 & $+0.030$ \\
    Corpus-graph backfill ($B{=}500$)    & 0.578 & 0.633 & $+0.056$ \\
    Dense @512, fused                    & 0.578 & 0.589 & $+0.011$ \\
    Dense @4096, fused                   & 0.578 & 0.649 & $+0.071$ \\
    \bottomrule
  \end{tabular}
\end{table}

\subsection{Reranking and Decision Levers}
\label{sec:posthoc_rerank}

Adding citation-graph features to the meta-learner improves F1 by 0.0141 on average across eight seeds, positive on every seed, at a paired $p=0.0078$.
The qualification matters more than the gain.
Computed naively, these features cost 0.18 F1, because a training query's own citation edge inflates the indegree of its gold candidates and produces a positive-class distribution the held-out queries do not share.
Recomputing the indegree leave-one-query-out aligns the two distributions and flips the sign.
The effect is small and real, and it is the only reranking-side change in this section that survives a seed-paired test.

Adding a feature that measures disagreement between the lexical and semantic scores raises the mean to 0.0164 above baseline.
That is 0.0023 above graph features alone, with two and a half times the variance and one seed in which it is negative.
We therefore do not claim the combined configuration improves on graph features alone.
The feature is redundant with the BM25 and cross-encoder scores the meta-learner already recombines, and a gradient-boosted model does not need help forming their ratio.

The decision rule offered the clearest opportunity and yielded nothing.
We implemented six alternatives to a global threshold: a score-gap rule, a top-ratio rule, a fixed $k$, a learned count regressor, a conformal rule, and an extreme-value cutoff.
All six lose, several of them heavily, as Table~\ref{tab:rerank_levers} shows.
This runs against the motivation for learned list truncation in legal search~\cite{lecut}, and we think the reason is specific rather than general.
Once the threshold is calibrated on an honest held-out slice, our score distribution is sharply peaked, and per-query rules mostly find new ways to be wrong about where its shoulder lies.
An oracle per-query $k$ would reach 0.385, so the headroom is real; no rule we tried captures it.

We also tried replacing our fine-tuned cross-encoder with a strong general-purpose reranker, given the same selectively truncated inputs.
Its NDCG@10 is 0.057 against our cross-encoder's 0.287, and it falls below BM25-RRF's 0.176 as well; through the meta-learner it costs 0.059 F1.
We state plainly what this does and does not show.
A zero-shot model losing to a domain-fine-tuned one is the expected result, not a discovery, and it is not evidence that dense reranking is unsuited to this task.
It shows only that no off-the-shelf substitute was available to us, and that improving this stage requires domain training rather than a swap.

Several COLIEE systems filter out candidates decided after the query case, which is sound in principle, since a case cannot cite its own future.
On our data the full filter costs 0.074 recall and 0.015 F1, and the selective variant is neutral at $+0.0002$.
The reason lies in the data rather than in the principle.
Our regex extraction dates 99.5\% of the corpus, but 607 of the 4,767 in-pool golds, 12.7\%, carry a date later than their query.
Some of those are extraction errors and some are genuine, because the years we recover are not always decision dates.
A reliable per-case decision date might make this lever work; regex-extracted years do not.

\begin{table}[t]
  \caption{Reranking and decision levers on the three-way split, wins and failures together. Only leakage-corrected graph features survive a seed-paired test. Oracle rows peek at test labels and bound the available headroom. Feature-set rows are means over eight seeds.}
  \label{tab:rerank_levers}
  \begin{tabular}{lcc}
    \toprule
    Configuration & F1 & $\Delta$F1 \\
    \midrule
    \multicolumn{3}{l}{\emph{Feature set} (global threshold, 8 seeds)} \\
    Base, 34 features                    & 0.3456 & -- \\
    $+$ graph features, leave-one-out    & 0.3596 & $+0.0141$ \\
    $+$ graph $+$ lexical/semantic gap   & 0.3619 & $+0.0164$ \\
    \midrule
    \multicolumn{3}{l}{\emph{Decision rule} (base features)} \\
    Global threshold                     & 0.3453 & -- \\
    Top-ratio                            & 0.3274 & $-0.0180$ \\
    Learned count regressor              & 0.3160 & $-0.0293$ \\
    Fixed $k$                            & 0.2933 & $-0.0520$ \\
    Extreme-value cutoff, best variant   & 0.2721 & $-0.0732$ \\
    Score-gap                            & 0.2629 & $-0.0825$ \\
    \emph{Oracle global threshold}       & 0.3523 & $+0.0070$ \\
    \emph{Oracle per-query $k$}          & 0.3847 & $+0.0394$ \\
    \midrule
    \multicolumn{3}{l}{\emph{Other}} \\
    Zero-shot reranker swap              & 0.2414 & $-0.0586$ \\
    Date filter, selective               & 0.3455 & $+0.0002$ \\
    Date filter, full                    & 0.3304 & $-0.0149$ \\
    \bottomrule
  \end{tabular}
\end{table}

\subsection{Evaluation Artifacts}
\label{sec:posthoc_artifacts}

Four results in this study reversed sign or magnitude once we corrected the protocol that produced them.
We report them because each was individually convincing at the time, and because three of the four inflated a conclusion we already wanted to reach.
Table~\ref{tab:artifacts} lists them.

Our first temporal experiment reported a calibration penalty of 0.188, which would have confirmed the workshop paper's diagnosis.
The threshold had been fit on out-of-fold scores from early-stopped models and applied to scores from a model trained on the whole training slice.
The two score scales differ, and the penalty measured that difference rather than any property of the data.
Fitting and applying with one model reduced it to 0.007.

Dense retrieval produced a recall of 0.043 in its first run, which we would have reported as a limitation of dense methods on legal text.
It was a usage bug.
The embedder pools at the final token, and our tokenisation appended no end-of-sequence token, so the pooled vector read a truncated content token instead.
Fixing that raised the same 50-query diagnostic from 0.046 to 0.506.
Even then the full run reached only 0.435, and only removing the 512-token truncation revealed the real figure of 0.611.
One bug and one window size had made a competitive method look useless.

The graph-feature reversal of Section~\ref{sec:posthoc_rerank} belongs on this list too, since $-0.18$ and $+0.014$ are the same feature under two protocols.
What made it detectable was not a better model but a distribution check.
Training golds had a citation indegree above zero in 100\% of cases against 37\% for held-out golds.
A feature whose positive class looks that different across the split is broken regardless of what the validation score says.

The fourth artifact is ours alone.
We had claimed a genuine test-set distribution shift, and we ran the embedding probe of Section~\ref{sec:posthoc_retest} expecting to quantify it.
It refuted the claim instead.
We record this because the asymmetry is the point: we had run no experiment capable of disconfirming that claim before we made it.

The four corrections share one discipline.
Fit and apply every parameter with the same model under a leave-one-out reference.
Compare the training and held-out feature distributions of the positive class before trusting a feature.
Reproduce on a cheap subset before scaling, which is how a 15-minute diagnostic over 50 queries replaced a repeat of a five-hour run.
None of this is sophisticated, and all of it is easier to skip when a result already agrees with expectation.

\begin{table*}[t]
  \caption{The four evaluation artifacts, the protocol fault behind each, the correction that exposed it, and the result before and after. \textbf{Each row reports a different metric and the desirable direction differs}, so the last column is not comparable down the table: the calibration penalty should be near zero, recall and $\Delta$F1 should be large and positive, and a Cohen's $d$ near zero means no distribution shift. The value after the arrow is the one reported elsewhere in this paper. Three of the four inflated a conclusion we already expected; the graph-feature artifact is the exception, and because it pointed against expectation it took a distribution check rather than a sanity check to find.}
  \label{tab:artifacts}
  \small
  \begin{tabular}{@{}l p{5.3cm} p{3.5cm} l r@{}}
    \toprule
    Artifact & Protocol fault & Correction & Metric & Naive $\rightarrow$ clean \\
    \midrule
    Score-scale mismatch
      & Threshold fit on out-of-fold scores from early-stopped models, then applied to a model trained on the whole slice
      & Fit and apply with one model
      & F1 penalty & $0.188 \rightarrow 0.007$ \\
    \addlinespace[3pt]
    Truncated document input
      & No end-of-sequence token for a last-token pooler, then a 512-token window on 2--5k-word judgments
      & Append EOS, embed the full document
      & recall@200 & $0.435 \rightarrow 0.611$ \\
    \addlinespace[3pt]
    Own-citation edge
      & A training query's own citation inflates the indegree of its gold candidates
      & Leave-one-query-out indegree
      & $\Delta$F1 & $-0.180 \rightarrow +0.014$ \\
    \addlinespace[3pt]
    Assumed test shift
      & Claimed without running an experiment that could have disconfirmed it
      & Label-free embedding probe
      & Cohen's $d$ & asserted $\rightarrow -0.16$ \\
    \bottomrule
  \end{tabular}
\end{table*}

\section{Limitations}
\label{sec:limitations}

The gold labels for the official test set were not released to participants.
The 43\% drop this paper studies therefore cannot be measured directly, and every experiment in Section~\ref{sec:posthoc} is a proxy assembled from the data we do hold.
When we write that a proposed cause does not reproduce, we mean it does not reproduce on any internal proxy we could construct.
That is weaker than a refutation and we do not want it read as one.

The leading untested suspect is the candidate corpus.
Official predictions were made against 1,848 documents; every number in Section~\ref{sec:posthoc} comes from the 9,556-document training corpus.
Those two collections differ in size by a factor of five and in composition in ways we cannot audit without the labels.
A retrieval system's precision depends on what else is in the pool, so a smaller and differently composed corpus could move F1 substantially on its own.
Our internal splits cannot reproduce that difference by construction, because they resample the same corpus.

The result is a narrower space of explanations rather than an answer.
We can say that three specific mechanisms do not account for the drop on the evidence available.
We cannot say what does.
We prefer to leave that stated plainly rather than promote the best surviving hypothesis into a conclusion the experiments do not support.

Section~\ref{sec:posthoc_errors} establishes that 637 missed golds reached the candidate pool and were ranked below the cut, and that a third of in-pool golds fall outside the cross-encoder's selective coverage.
It does not establish which of those facts causes the other.
Expanding coverage would help if the cross-encoder would score these golds highly once shown them, and would not if the model is simply wrong about them.
We ran a full-coverage experiment intended to separate the two, but it did not leave a reproducible artifact, so we exclude its result rather than cite a number we cannot regenerate.
This is the largest open question the paper leaves.

The distribution probe has a narrow scope.
It compares query documents, so it speaks to query text and to nothing else.
A shift in the candidate corpus, in the citation label distribution, or in a property neither embedding encodes would be invisible to it.
Running it in two embedding spaces removes the single-encoder objection but not this one.

The recall levers of Section~\ref{sec:posthoc_recall} were each measured alone.
We report no combined figure because we ran no combined experiment, and we would rather understate the ceiling than estimate it.
Two further costs are unmeasured.
Enlarging the pool hands the reranker more candidates to be wrong about, and the corpus-graph lever in particular increases the pool by a factor of 2.5 without our having checked what that does to precision.

Everything here comes from one task, one corpus, and one pipeline.
The negative results are the ones most at risk of being over-generalised, so we state their scope explicitly: per-query cutoffs lost to a global threshold on our score distribution, and a date filter failed given our regex-extracted years.
Neither is a claim about legal case retrieval in general.
The seed-paired protocol of Section~\ref{sec:posthoc_protocol} makes our own comparisons more reliable, and it says nothing about whether they transfer.

\section{Conclusion}

We proposed a four-stage hybrid pipeline for legal case retrieval that combines lexical retrieval, neural reranking, graph-based community features, and gradient-boosted meta-learning.
The ablation study confirms the cross-encoder with selective truncation as the most impactful component ($-$24\% relative F1 when removed), while graph-based features from GraphRAG Lite and the GNN reranker provide modest but consistent gains ($-$2.0\% and $-$3.8\%, respectively).

Our system achieved F1=0.177 on the COLIEE 2026 test set, with our best run (run~2, recall-favouring) ranking 35th of 54 official runs and 15th of the 22 participating teams by each team's best run.
That is well below our cross-validated estimate of 0.311, and we proposed three causes for the difference.
Section~\ref{sec:posthoc} tested all three and none of them survives.
Threshold transfer costs 0.007 F1 rather than the double-digit penalty we assumed, decision quality is flat across four chronological quartiles, and the official test queries are no farther from the training manifold than the training queries are from each other, in two embedding spaces.
We can rule these mechanisms out on every internal proxy available to us; we cannot say what replaces them, and Section~\ref{sec:limitations} says so.

What the measurements do support is a different division of effort.
The misses split exactly evenly between candidates that never reached the pool and candidates that reached it and were ranked below the cut, so the recall ceiling accounts for half the damage rather than most of it.
The recall half is also the cheaper half.
Tuning BM25 length normalisation, adding an event-triple view, and embedding full documents rather than truncated ones each lift top-200 recall by three to seven points, and two of the three need no GPU at all.
The reranking half resisted every change we tried except leakage-corrected graph features, which add 0.014 F1 across eight seeds.

Three directions follow from this.
The recall levers should be combined and measured together, including the precision cost of the larger pool, which we did not test.
The reranking stage needs a model trained on this domain rather than a substitution, since an off-the-shelf zero-shot reranker performs worse than BM25 here.
Most importantly, someone should settle whether the in-pool misses are a coverage failure or a quality failure.
A third of the golds that retrieval surfaces are never scored by the cross-encoder at all.
Which of those two answers holds decides where the next effort belongs.

We close on the part we did not anticipate when we started.
Four separate results in this study reversed once we corrected the protocol that produced them, and three of the four had inflated a conclusion we already expected.
The checks that caught them are cheap: fit and apply parameters with one model, compare the positive-class feature distribution across the split before trusting a feature, and reproduce on a small subset before committing to a long run.
We report them in full because a result that agrees with expectation is the one least likely to be audited, and on this evidence it is the one most likely to need it.

\begin{acks}
This work was supported by the IITP(Institute of Information \& Communications Technology Planning \& Evaluation)-ITRC(Information Technology Research Center) grant funded by the Korea government(Ministry of Science and ICT)(IITP-2026-RS-2024-00436936).
This research was supported by the MSIT(Ministry of Science and ICT), Korea, under the ICAN(ICT Challenge and Advanced Network of HRD) support program(IITP-2026-RS-2023-00259497) supervised by the IITP(Institute for Information \& Communications Technology Planning \& Evaluation).
This research was supported by the MSIT(Ministry of Science and ICT), Korea, under the Graduate School of Virtual Convergence support program(IITP-2026-RS-2023-00254129) supervised by the IITP(Institute for Information \& Communications Technology Planning \& Evaluation).
\end{acks}

\bibliographystyle{ACM-Reference-Format}
\bibliography{references}


\begin{thebibliography}{25}


\ifx \showCODEN    \undefined \def \showCODEN     #1{\unskip}     \fi
\ifx \showDOI      \undefined \def \showDOI       #1{#1}\fi
\ifx \showISBNx    \undefined \def \showISBNx     #1{\unskip}     \fi
\ifx \showISBNxiii \undefined \def \showISBNxiii  #1{\unskip}     \fi
\ifx \showISSN     \undefined \def \showISSN      #1{\unskip}     \fi
\ifx \showLCCN     \undefined \def \showLCCN      #1{\unskip}     \fi
\ifx \shownote     \undefined \def \shownote      #1{#1}          \fi
\ifx \showarticletitle \undefined \def \showarticletitle #1{#1}   \fi
\ifx \showURL      \undefined \def \showURL       {\relax}        \fi
\providecommand\bibfield[2]{#2}
\providecommand\bibinfo[2]{#2}
\providecommand\natexlab[1]{#1}
\providecommand\showeprint[2][]{arXiv:#2}

\bibitem[Chen et~al\mbox{.}(2024)]%
        {bgem3}
\bibfield{author}{\bibinfo{person}{Jianlyu Chen}, \bibinfo{person}{Shitao
  Xiao}, \bibinfo{person}{Peitian Zhang}, \bibinfo{person}{Kun Luo},
  \bibinfo{person}{Defu Lian}, {and} \bibinfo{person}{Zheng Liu}.}
  \bibinfo{year}{2024}\natexlab{}.
\newblock \showarticletitle{M3-Embedding: Multi-Linguality,
  Multi-Functionality, Multi-Granularity Text Embeddings Through Self-Knowledge
  Distillation}. In \bibinfo{booktitle}{\emph{Findings of the Association for
  Computational Linguistics: ACL 2024}}. \bibinfo{pages}{2318--2335}.
\newblock


\bibitem[Cormack et~al\mbox{.}(2009)]%
        {rrf}
\bibfield{author}{\bibinfo{person}{Gordon~V Cormack},
  \bibinfo{person}{Charles~LA Clarke}, {and} \bibinfo{person}{Stefan
  Buettcher}.} \bibinfo{year}{2009}\natexlab{}.
\newblock \showarticletitle{Reciprocal rank fusion outperforms condorcet and
  individual rank learning methods}. In \bibinfo{booktitle}{\emph{Proceedings
  of the 32nd international ACM SIGIR conference on Research and development in
  information retrieval}}. \bibinfo{pages}{758--759}.
\newblock


\bibitem[Dai and Callan(2020)]%
        {dai2020context}
\bibfield{author}{\bibinfo{person}{Zhuyun Dai} {and} \bibinfo{person}{Jamie
  Callan}.} \bibinfo{year}{2020}\natexlab{}.
\newblock \showarticletitle{Context-Aware Term Weighting For First Stage
  Passage Retrieval}. In \bibinfo{booktitle}{\emph{Proceedings of the 43rd
  International ACM SIGIR Conference on Research and Development in Information
  Retrieval}}. \bibinfo{pages}{1533--1536}.
\newblock


\bibitem[Feng et~al\mbox{.}(2024)]%
        {feng2024survey}
\bibfield{author}{\bibinfo{person}{Yi Feng}, \bibinfo{person}{Chuanyi Li},
  {and} \bibinfo{person}{Vincent Ng}.} \bibinfo{year}{2024}\natexlab{}.
\newblock \showarticletitle{Legal case retrieval: A survey of the state of the
  art}. In \bibinfo{booktitle}{\emph{Proceedings of the 62nd Annual Meeting of
  the Association for Computational Linguistics (Volume 1: Long Papers)}}.
  \bibinfo{pages}{6472--6485}.
\newblock


\bibitem[Goebel et~al\mbox{.}(2025)]%
        {coliee2025}
\bibfield{author}{\bibinfo{person}{Randy Goebel}, \bibinfo{person}{Yoshinobu
  Kano}, \bibinfo{person}{Mi-Young Kim}, \bibinfo{person}{Calum Kwan},
  \bibinfo{person}{Ken Satoh}, \bibinfo{person}{Hiroaki Yamada}, {and}
  \bibinfo{person}{Masaharu Yoshioka}.} \bibinfo{year}{2025}\natexlab{}.
\newblock \showarticletitle{An Overview of the COLIEE 2025 Competition: Legal
  Case Law and Statute Law Information Retrieval and Entailment}. In
  \bibinfo{booktitle}{\emph{Proceedings of the Twentieth International
  Conference on Artificial Intelligence and Law}}. \bibinfo{pages}{506--515}.
\newblock


\bibitem[Goebel et~al\mbox{.}(2023)]%
        {coliee2023}
\bibfield{author}{\bibinfo{person}{Randy Goebel}, \bibinfo{person}{Yoshinobu
  Kano}, \bibinfo{person}{Mi-Young Kim}, \bibinfo{person}{Juliano Rabelo},
  \bibinfo{person}{Ken Satoh}, {and} \bibinfo{person}{Masaharu Yoshioka}.}
  \bibinfo{year}{2023}\natexlab{}.
\newblock \showarticletitle{Summary of the competition on legal information,
  extraction/entailment (COLIEE) 2023}. In
  \bibinfo{booktitle}{\emph{Proceedings of the nineteenth international
  conference on artificial intelligence and law}}. \bibinfo{pages}{472--480}.
\newblock


\bibitem[He et~al\mbox{.}(2021)]%
        {deberta}
\bibfield{author}{\bibinfo{person}{Pengcheng He}, \bibinfo{person}{Xiaodong
  Liu}, \bibinfo{person}{Jianfeng Gao}, {and} \bibinfo{person}{Weizhu Chen}.}
  \bibinfo{year}{2021}\natexlab{}.
\newblock \showarticletitle{{DeBERTa}: Decoding-enhanced {BERT} with
  Disentangled Attention}. In \bibinfo{booktitle}{\emph{International
  Conference on Learning Representations (ICLR)}}.
\newblock
\urldef\tempurl%
\url{https://openreview.net/forum?id=XPZIaotutsD}
\showURL{%
\tempurl}


\bibitem[Hu et~al\mbox{.}(2022)]%
        {hulora}
\bibfield{author}{\bibinfo{person}{Edward~J Hu}, \bibinfo{person}{Yelong Shen},
  \bibinfo{person}{Phillip Wallis}, \bibinfo{person}{Zeyuan Allen-Zhu},
  \bibinfo{person}{Yuanzhi Li}, \bibinfo{person}{Shean Wang},
  \bibinfo{person}{Lu Wang}, {and} \bibinfo{person}{Weizhu Chen}.}
  \bibinfo{year}{2022}\natexlab{}.
\newblock \showarticletitle{LoRA: Low-Rank Adaptation of Large Language
  Models}. In \bibinfo{booktitle}{\emph{International Conference on Learning
  Representations}}.
\newblock


\bibitem[Joshi et~al\mbox{.}(2023)]%
        {joshi2023u}
\bibfield{author}{\bibinfo{person}{Abhinav Joshi}, \bibinfo{person}{Akshat
  Sharma}, \bibinfo{person}{Sai~Kiran Tanikella}, {and}
  \bibinfo{person}{Ashutosh Modi}.} \bibinfo{year}{2023}\natexlab{}.
\newblock \showarticletitle{{U-CREAT:} Unsupervised Case Retrieval using Events
  extrAcTion}. In \bibinfo{booktitle}{\emph{Proceedings of the 61st Annual
  Meeting of the Association for Computational Linguistics (Volume 1: Long
  Papers), {ACL} 2023, Toronto, Canada, July 9-14, 2023}},
  \bibfield{editor}{\bibinfo{person}{Anna Rogers}, \bibinfo{person}{Jordan~L.
  Boyd{-}Graber}, {and} \bibinfo{person}{Naoaki Okazaki}} (Eds.).
  \bibinfo{publisher}{Association for Computational Linguistics},
  \bibinfo{pages}{13899--13915}.
\newblock
\urldef\tempurl%
\url{https://doi.org/10.18653/V1/2023.ACL-LONG.777}
\showDOI{\tempurl}


\bibitem[Ke et~al\mbox{.}(2017)]%
        {lgbm}
\bibfield{author}{\bibinfo{person}{Guolin Ke}, \bibinfo{person}{Qi Meng},
  \bibinfo{person}{Thomas Finley}, \bibinfo{person}{Taifeng Wang},
  \bibinfo{person}{Wei Chen}, \bibinfo{person}{Weidong Ma},
  \bibinfo{person}{Qiwei Ye}, {and} \bibinfo{person}{Tie-Yan Liu}.}
  \bibinfo{year}{2017}\natexlab{}.
\newblock \showarticletitle{Lightgbm: A highly efficient gradient boosting
  decision tree}.
\newblock \bibinfo{journal}{\emph{Advances in neural information processing
  systems}}  \bibinfo{volume}{30} (\bibinfo{year}{2017}).
\newblock


\bibitem[Ko et~al\mbox{.}(2025)]%
        {ko2025when}
\bibfield{author}{\bibinfo{person}{Dayoon Ko}, \bibinfo{person}{Jinyoung Kim},
  \bibinfo{person}{Sohyeon Kim}, \bibinfo{person}{Jinhyuk Kim},
  \bibinfo{person}{Jaehoon Lee}, \bibinfo{person}{Seonghak Song},
  \bibinfo{person}{Minyoung Lee}, {and} \bibinfo{person}{Gunhee Kim}.}
  \bibinfo{year}{2025}\natexlab{}.
\newblock \showarticletitle{When Should Dense Retrievers Be Updated in Evolving
  Corpora? Detecting Out-of-Distribution Corpora Using GradNormIR}. In
  \bibinfo{booktitle}{\emph{Findings of the Association for Computational
  Linguistics, {ACL} 2025, Vienna, Austria, July 27 - August 1, 2025}}
  \emph{(\bibinfo{series}{Findings of {ACL}}, Vol.~\bibinfo{volume}{{ACL}
  2025})}, \bibfield{editor}{\bibinfo{person}{Wanxiang Che},
  \bibinfo{person}{Joyce Nabende}, \bibinfo{person}{Ekaterina Shutova}, {and}
  \bibinfo{person}{Mohammad~Taher Pilehvar}} (Eds.).
  \bibinfo{publisher}{Association for Computational Linguistics},
  \bibinfo{pages}{25977--25996}.
\newblock
\urldef\tempurl%
\url{https://doi.org/10.18653/V1/2025.FINDINGS-ACL.1334}
\showDOI{\tempurl}


\bibitem[Li et~al\mbox{.}(2023)]%
        {sailer}
\bibfield{author}{\bibinfo{person}{Haitao Li}, \bibinfo{person}{Qingyao Ai},
  \bibinfo{person}{Jia Chen}, \bibinfo{person}{Qian Dong},
  \bibinfo{person}{Yueyue Wu}, \bibinfo{person}{Yiqun Liu},
  \bibinfo{person}{Chong Chen}, {and} \bibinfo{person}{Qi Tian}.}
  \bibinfo{year}{2023}\natexlab{}.
\newblock \showarticletitle{Sailer: structure-aware pre-trained language model
  for legal case retrieval}. In \bibinfo{booktitle}{\emph{Proceedings of the
  46th International ACM SIGIR Conference on Research and Development in
  Information Retrieval}}. \bibinfo{pages}{1035--1044}.
\newblock


\bibitem[Li et~al\mbox{.}(2024)]%
        {li2024towards}
\bibfield{author}{\bibinfo{person}{Haitao Li}, \bibinfo{person}{You Chen},
  \bibinfo{person}{Zhekai Ge}, \bibinfo{person}{Qingyao Ai},
  \bibinfo{person}{Yiqun Liu}, \bibinfo{person}{Quan Zhou}, {and}
  \bibinfo{person}{Shuai Huo}.} \bibinfo{year}{2024}\natexlab{}.
\newblock \showarticletitle{Towards an in-depth comprehension of case relevance
  for better legal retrieval}. In \bibinfo{booktitle}{\emph{JSAI International
  Symposium on Artificial Intelligence}}. Springer, \bibinfo{pages}{212--227}.
\newblock


\bibitem[Louis et~al\mbox{.}(2023)]%
        {louis2023}
\bibfield{author}{\bibinfo{person}{Antoine Louis}, \bibinfo{person}{Gijs
  Van~Dijck}, {and} \bibinfo{person}{Gerasimos Spanakis}.}
  \bibinfo{year}{2023}\natexlab{}.
\newblock \showarticletitle{Finding the law: Enhancing statutory article
  retrieval via graph neural networks}. In
  \bibinfo{booktitle}{\emph{Proceedings of the 17th Conference of the European
  Chapter of the Association for Computational Linguistics}}.
  \bibinfo{pages}{2761--2776}.
\newblock


\bibitem[Ma et~al\mbox{.}(2022)]%
        {lecut}
\bibfield{author}{\bibinfo{person}{Yixiao Ma}, \bibinfo{person}{Qingyao Ai},
  \bibinfo{person}{Yueyue Wu}, \bibinfo{person}{Yunqiu Shao},
  \bibinfo{person}{Yiqun Liu}, \bibinfo{person}{Min Zhang}, {and}
  \bibinfo{person}{Shaoping Ma}.} \bibinfo{year}{2022}\natexlab{}.
\newblock \showarticletitle{Incorporating retrieval information into the
  truncation of ranking lists for better legal search}. In
  \bibinfo{booktitle}{\emph{Proceedings of the 45th International ACM SIGIR
  Conference on Research and Development in Information Retrieval}}.
  \bibinfo{pages}{438--448}.
\newblock


\bibitem[Ma et~al\mbox{.}(2023)]%
        {slr}
\bibfield{author}{\bibinfo{person}{Yixiao Ma}, \bibinfo{person}{Yueyue Wu},
  \bibinfo{person}{Qingyao Ai}, \bibinfo{person}{Yiqun Liu},
  \bibinfo{person}{Yunqiu Shao}, \bibinfo{person}{Min Zhang}, {and}
  \bibinfo{person}{Shaoping Ma}.} \bibinfo{year}{2023}\natexlab{}.
\newblock \showarticletitle{Incorporating structural information into legal
  case retrieval}.
\newblock \bibinfo{journal}{\emph{ACM Transactions on Information Systems}}
  \bibinfo{volume}{42}, \bibinfo{number}{2} (\bibinfo{year}{2023}),
  \bibinfo{pages}{1--28}.
\newblock


\bibitem[MacAvaney et~al\mbox{.}(2022)]%
        {macavaney2022adaptive}
\bibfield{author}{\bibinfo{person}{Sean MacAvaney}, \bibinfo{person}{Nicola
  Tonellotto}, {and} \bibinfo{person}{Craig Macdonald}.}
  \bibinfo{year}{2022}\natexlab{}.
\newblock \showarticletitle{Adaptive Re-Ranking with a Corpus Graph}. In
  \bibinfo{booktitle}{\emph{Proceedings of the 31st {ACM} International
  Conference on Information {\&} Knowledge Management, Atlanta, GA, USA,
  October 17-21, 2022}}, \bibfield{editor}{\bibinfo{person}{Mohammad~Al Hasan}
  {and} \bibinfo{person}{Li~Xiong}} (Eds.). \bibinfo{publisher}{{ACM}},
  \bibinfo{pages}{1491--1500}.
\newblock
\urldef\tempurl%
\url{https://doi.org/10.1145/3511808.3557231}
\showDOI{\tempurl}


\bibitem[Nguyen et~al\mbox{.}(2025)]%
        {jnlp2025}
\bibfield{author}{\bibinfo{person}{Hai Nguyen} {et~al\mbox{.}}}
  \bibinfo{year}{2025}\natexlab{}.
\newblock \showarticletitle{JNLP at COLIEE 2025: Hybrid Large Language
  Model-based Framework for Legal Information Retrieval and Entailment}. In
  \bibinfo{booktitle}{\emph{Proceedings of the Workshop on the Twelfth
  International Competition on Legal Information Extraction and Entailment
  COLIEE 2025 in association with the 20th International Conference on
  Artificial Intelligence and Law}}, \bibfield{editor}{\bibinfo{person}{Randy
  Goebel} {et~al\mbox{.}}} (Eds.).
\newblock


\bibitem[Paul et~al\mbox{.}(2022)]%
        {lesicin}
\bibfield{author}{\bibinfo{person}{Shounak Paul}, \bibinfo{person}{Pawan
  Goyal}, {and} \bibinfo{person}{Saptarshi Ghosh}.}
  \bibinfo{year}{2022}\natexlab{}.
\newblock \showarticletitle{Lesicin: A heterogeneous graph-based approach for
  automatic legal statute identification from indian legal documents}. In
  \bibinfo{booktitle}{\emph{Proceedings of the AAAI conference on artificial
  intelligence}}, Vol.~\bibinfo{volume}{36}. \bibinfo{pages}{11139--11146}.
\newblock


\bibitem[Peng et~al\mbox{.}(2025)]%
        {graphrag_survey}
\bibfield{author}{\bibinfo{person}{Boci Peng}, \bibinfo{person}{Yun Zhu},
  \bibinfo{person}{Yongchao Liu}, \bibinfo{person}{Xiaohe Bo},
  \bibinfo{person}{Haizhou Shi}, \bibinfo{person}{Chuntao Hong},
  \bibinfo{person}{Yan Zhang}, {and} \bibinfo{person}{Siliang Tang}.}
  \bibinfo{year}{2025}\natexlab{}.
\newblock \showarticletitle{Graph retrieval-augmented generation: A survey}.
\newblock \bibinfo{journal}{\emph{ACM Transactions on Information Systems}}
  \bibinfo{volume}{44}, \bibinfo{number}{2} (\bibinfo{year}{2025}),
  \bibinfo{pages}{1--52}.
\newblock


\bibitem[Traag et~al\mbox{.}(2019)]%
        {leiden}
\bibfield{author}{\bibinfo{person}{Vincent~A Traag}, \bibinfo{person}{Ludo
  Waltman}, {and} \bibinfo{person}{Nees~Jan Van~Eck}.}
  \bibinfo{year}{2019}\natexlab{}.
\newblock \showarticletitle{From Louvain to Leiden: guaranteeing well-connected
  communities}.
\newblock \bibinfo{journal}{\emph{Scientific reports}} \bibinfo{volume}{9},
  \bibinfo{number}{1} (\bibinfo{year}{2019}), \bibinfo{pages}{5233}.
\newblock


\bibitem[Tran et~al\mbox{.}(2019)]%
        {tran2019}
\bibfield{author}{\bibinfo{person}{Vu Tran}, \bibinfo{person}{Minh~Le Nguyen},
  {and} \bibinfo{person}{Ken Satoh}.} \bibinfo{year}{2019}\natexlab{}.
\newblock \showarticletitle{Building legal case retrieval systems with lexical
  matching and summarization using a pre-trained phrase scoring model}. In
  \bibinfo{booktitle}{\emph{Proceedings of the seventeenth international
  conference on artificial intelligence and law}}. \bibinfo{pages}{275--282}.
\newblock


\bibitem[Veli{\v{c}}kovi{\'c} et~al\mbox{.}(2018)]%
        {gat}
\bibfield{author}{\bibinfo{person}{Petar Veli{\v{c}}kovi{\'c}},
  \bibinfo{person}{Guillem Cucurull}, \bibinfo{person}{Arantxa Casanova},
  \bibinfo{person}{Adriana Romero}, \bibinfo{person}{Pietro Li{\`o}}, {and}
  \bibinfo{person}{Yoshua Bengio}.} \bibinfo{year}{2018}\natexlab{}.
\newblock \showarticletitle{Graph Attention Networks}. In
  \bibinfo{booktitle}{\emph{International Conference on Learning
  Representations}}.
\newblock


\bibitem[Xiao et~al\mbox{.}(2024)]%
        {bge}
\bibfield{author}{\bibinfo{person}{Shitao Xiao}, \bibinfo{person}{Zheng Liu},
  \bibinfo{person}{Peitian Zhang}, \bibinfo{person}{Niklas Muennighoff},
  \bibinfo{person}{Defu Lian}, {and} \bibinfo{person}{Jian-Yun Nie}.}
  \bibinfo{year}{2024}\natexlab{}.
\newblock \showarticletitle{C-pack: Packed resources for general chinese
  embeddings}. In \bibinfo{booktitle}{\emph{Proceedings of the 47th
  international ACM SIGIR conference on research and development in information
  retrieval}}. \bibinfo{pages}{641--649}.
\newblock


\bibitem[Zhang et~al\mbox{.}(2023)]%
        {cfgl}
\bibfield{author}{\bibinfo{person}{Kun Zhang}, \bibinfo{person}{Chong Chen},
  \bibinfo{person}{Yuanzhuo Wang}, \bibinfo{person}{Qi Tian}, {and}
  \bibinfo{person}{Long Bai}.} \bibinfo{year}{2023}\natexlab{}.
\newblock \showarticletitle{Cfgl-lcr: A counterfactual graph learning framework
  for legal case retrieval}. In \bibinfo{booktitle}{\emph{Proceedings of the
  29th ACM SIGKDD Conference on knowledge discovery and data mining}}.
  \bibinfo{pages}{3332--3341}.
\newblock


\end{thebibliography}

\end{document}